\documentclass[english]{article}
\usepackage[T1]{fontenc}
\usepackage[latin9]{inputenc}
\usepackage[a4paper]{geometry}
\usepackage{babel}
\usepackage{url}
\usepackage{amsmath}
\usepackage{graphicx}
\usepackage{setspace}
\usepackage[unicode=true,
 bookmarks=true,bookmarksnumbered=false,bookmarksopen=false,
 breaklinks=false,pdfborder={0 0 1},backref=false,colorlinks=false]
 {hyperref}

\makeatletter

\providecommand{\tabularnewline}{\\}

\usepackage{cite}

\date{}

\DeclareMathOperator{\rmd}{d}

\makeatother

\begin{document}

\title{Empirical analysis and agent-based modeling of the Lithuanian parliamentary
elections}

\author{Aleksejus Kononovicius}

\date{Vilnius University, Institute of Theoretical Physics and Astronomy}
\maketitle
\begin{abstract}
In this contribution we analyze a parties' vote share distribution
across the polling stations during the Lithuanian parliamentary elections
of 1992, 2008 and 2012. We find that the distribution is rather well
fitted by the Beta distribution. To reproduce this empirical observation
we propose a simple multi-state agent-based model of the voting behavior.
In the proposed model agents change the party they vote for either
idiosyncratically or due to a linear recruitment mechanism. We use
the model to reproduce the vote share distribution observed during
the election of 1992. We discuss model extensions needed to reproduce
the vote share distribution observed during the other elections.
\end{abstract}

\section{Introduction}

While any individual vote is equally important to determine the outcome
of an election, the probability for a single vote to decide the outcome
is extremely small. As utility of casting a vote in this context seems
to be small and as there is at least a minor associated cost, it seems
that a rational choice would be simply not to vote. Yet this simplistic
context may be further extended to provide a sound reasoning for why
people vote. Some argue that people vote to show a support for the
political system \cite{Riker1968} or to avoid a risk of regret \cite{Ferejohn1974},
there might also be a social cost for abstension \cite{Uhlaner1989}.
Some of the aforementioned works as well as numerous other earlier
game theoretic approaches, such as \cite{Hotelling1929,Downs1957,Black1958},
had shown promise that game theoretic voting models would soon provide
rich and sophisticated explanation for the voting behavior. Yet further
research have shown that general game theoretic models of the voting
behavior with pure Nash equilibrium, and even mixed Nash equilibrium,
might be impossible unless under certain specific conditions \cite{Plott1967,McKelvey1976,Ansolabehere2000}.
But people are rarely well informed and ideally rational, as they
are not homo economicus nor are they Laplace's demons, \cite{Arthur1994AER,Akerlof2009Princeton,Helbing2013Kirman}.

The above context provides a good reasoning to consider the modeling
of the voting behavior from the perspective of psychology \cite{Nowak1990,Axelrod1997JConfRes,Hegselmann2002JASSS,Deffuant2006JASSS,Sobkowicz2012PlosOne,Lilleker2014,Sobkowicz2016PlosOne,Duggins2017JASSS}.
The main drawback of these psychologically motivated models is that
they usually are rather complicated, at least when compared with game
theoretic models, hard to implement and understand the obtained results.
Also usually these models involve a large number of parameters, which
may lead to overfitting the data or different parameter sets providing
similar results. Notably recently a psychologically motivated model
was successfully used to predict the Polish election of 2015 \cite{Sobkowicz2012PlosOne,Sobkowicz2016PlosOne}.

Another possible approach to the modeling of the voting behavior has
its roots in statistical physics. This perspective could be neatly
summarized by quoting the Boltzmann's molecular chaos hypothesis \cite{Ball2002PhysA}:
\begin{quotation}
The molecules are like so many individuals, having the most various
states of motion, and the properties of gases only remain unaltered
because the number of these molecules which on the average have a
given state of motion is constant.
\end{quotation}
During the last three decades physicists have approached social and
economic systems from this perspective, looking for universal laws
and important statistical patterns, while proposing simple theoretical
models to explain the empirical observations. This effort by a numerous
more or less prominent physicists became what is now known as sociophysics
and econophysics \cite{Galam1986JMatPsych,Mantegna2000Cambridge,Chakrabarti2006Wiley,Castellano2009RevModPhys,Stauffer2013JStatPhys,Abergel2017Springer}.
The opinion dynamics, and the voting behavior as a proxy of opinion,
is still one of the major topics in sociophysics \cite{SznajdWeron2005APPB,Galam2008ModPhysC,Castellano2009RevModPhys,Stauffer2013JStatPhys,Nyczka2013,Abergel2017Springer}.

This paper contributes to the understanding and describtion of the
voting behavior from a couple of different point of views. First of
all Lithuania is a young democratic nation and the analysis of the
Lithuanian parliamentary elections' data sets seems interesting in
the context of similar analyses carried out on the data sets gathered
in the mature democratic nations, such as Brazil, England, Germany,
France, Finland, Norway or Switzerland \cite{Fortunato2007PRL,Andersen2008IntJModPhysC,Borghesi2010EPJB,Mantovani2011EPL,Borghesi2012PLOS}.
In the political science and sociological literature one would find
numerous previous approaches to the Lithuanian parliamentary elections,
e.g., \cite{Krupavicius1997EStud,Degutis2000LitPolSci,Kreuzer2003StudCompIntDev,Jurkynas2004JBaltStud,Ramonaite2006}.
Yet most of these approaches had a quite different perspective, most
of these papers discuss general electoral trends in the context of
social, demographic and economic changes. For this kind of discussion
a highly aggregated (e.g., on a municipal district level) data sets
prove to be sufficient, while in this paper we will consider the data
on the smallest scale available (polling station level).

Another key contribution of this paper is a simple agent-based model,
which is used to explain the statistical patterns uncovered during
the empirical analysis. The proposed model is built upon a two-state
herding model originally proposed by Kirman in \cite{Kirman1993QJE}.
In the recent years the two-state herding model was quite frequently
and rather successfully applied to reproduce the statistical patterns
observed in the empirical data of the financial markets \cite{Alfarano2005CompEco,Alfarano2008Dyncon,Alfi2009EPJB1,Kononovicius2012PhysA,Alfarano2013EJF,Gontis2014PlosOne,Gontis2016PhysA}.
In this paper we extend the two-state herding model to allow the agents
to switch between more than two states. We discuss the similarity
between the proposed model and the well known Voter model \cite{Clifford1973,Liggett1999,Galam2002PhysA,Castellano2003EPL,FernandezGarcia2014PRL,Sano2017}.

Our approach is unique in a sense that we consider reproducing the
parties' vote share distribution observed in the Lithuanian parliamentary
elections. In the previous literature there was only a single attempt
to model, and predict, popular vote (aggregated vote share) in the
Lithuanian parliamentary elections using regression model, see \cite{Jastramskis2011PS}.
Numerous previous sociophysics papers have mainly ignored the vote
share distribution, likely due to belief that the vote share distribution
reflects electoral sensitivity to the policies promoted by the parties
and less to the endogenous interactions between the voters (a similar
argument is given in \cite{Fortunato2007PRL}). To some extent this
belief is supported by game theoretic models, see \cite{Downs1957,Black1958}.
Notably there were a couple of sociophysics papers considering two-state
agent-based models of the voting behavior, e.g., voting for or against
certain proposals in a referendum \cite{Galam2008ModPhysC}. It is
interesting to note that recently the binary models considered in
\cite{Galam2008ModPhysC} were used to construct a simple financial
market model \cite{Galam2016Chaos}, while we start from the financial
market model\cite{Gontis2014PlosOne} and move towards the model of
the voting behavior. While the vote share distribution was mainly
ignored in the previous sociophysics papers, the other statistical
patterns arising during the many different elections were considered
for the empirical analysis and modeling: a branching process model
was proposed to reproduce the individual politician, nominated via
open party list, vote share distribution \cite{Fortunato2007PRL},
a network model was used to explain how people decide whether to take
part in the municipal elections \cite{Mantovani2011EPL}, a diffusive
model for the turn-out was proposed in \cite{Borghesi2010EPJB,Borghesi2012PLOS}.
One of a more similar approaches was taken by \cite{Fenner2016},
in which a generative model was proposed to reproduce the rank-size
distribution of parties' vote share. Another similar approach, taken
by \cite{Sano2017}, considered the vote share distribution observed
in the elections of House of Representatives in Japan. Latter approach
\cite{Sano2017} also used a mean-field Voter model to explain the
empirical observations.

This paper is organized as follows. In Section~\ref{sec:empirical-analysis}
we discuss the Lithuanian parliamentary election system as well as
carry out the empirical analysis. Next, in Section~\ref{sec:abm},
we briefly introduce the two-state herding model and extend it to
account for the multiple states. Afterwards, in Section~\ref{sec:apply-abm},
we apply the extended model to reproduce the statistical patterns
uncovered during the empirical analysis. Finally we end the paper
with a discussion (see Section~\ref{sec:conclusions}).

\section{Empirical analysis of the data from the Lithuanian parliamentary
elections\label{sec:empirical-analysis}}

Let us start by discussing the parliamentary voting system used in
Lithuania. Lithuanian parliamentary elections are held every $4$
years. During every election all of the $141$ parliamentary seats
are distributed using two-tier voting system. Namely, $71$ seats
in the parliament are taken by elected district representatives (there
are $71$ electoral districts in total), while the other $70$ seats
are distributed according to the popular vote among the parties that
received more than $5\%$ of the popular vote. In other words, each
individual voter is able to vote for a single candidate to represent
his electoral district (the two round voting system is used) and for
an open party list (listing up to $5$ individuals from that list).
Every electoral district has multiple polling stations (their number
varies over the years), which further subdivide the electoral districts.
Every eligible voter is assigned to a single polling station based
on location of their residence. Each of the polling stations may have
widely different number of the assigned voters \textendash{} some
of the smallest polling stations have as few as $100$ assigned voters,
while the largest have up to $7000$ assigned voters.

In this paper we consider only votes cast for the open party lists
in each of the local polling stations. We do not analyze ranking of
the individuals on the party lists (similar analysis was carried out
in, e.g., \cite{Fortunato2007PRL}), voting for the representative
of electoral district (similar data was previously considered in,
e.g., \cite{Fenner2016,Sano2017}) nor turnout rates (modeling and
analysis of which was previously considered in, e.g., \cite{Borghesi2010EPJB,Borghesi2012PLOS}).
We ignore votes cast in the polling stations abroad or votes cast
by post. In the analysis that follows we consider only parties that
were elected to the parliament (total vote share larger thant $5\%$),
while all other less succesful parties were combined into a single
party, which we have labeled as the ``Other'' party.

In this paper we consider the three data sets from the Lithuanian
parliamentary elections of 1992, 2008 and 2012. All of the original
data sets were made publicly available by the Central Electoral Commission
of the Republic of Lithuania (at \url{https://www.rinkejopuslapis.lt/ataskaitu-formavimas}).
We have downloaded the original data sets from the website on August
31, 2016. During the preliminary phase of the empirical analysis we
have found some small inconsistencies within the original data. The
original 1992 election data set had seven polling stations with incorrect
total vote counts. We have identified three pairs of polling stations
which were, most likely, swapped among themselves as the number of
missing votes in the one polling station matched the number of surplus
votes in the other. While we have dealt with the remaining polling
station by simply adjusting the total vote count to match the sum
of votes cast for each of the parties in that polling station. We
have also found that data from $51$ (out of $2034$) polling stations
is missing (the data was filled with zeros) from the original 2008
election data set. We have not identified any issues with the original
2012 election data set. These minor inconsistencies would not impact
the overall result of any of the considered elections nor the results
reported in this section. We have made the modified data sets available
online at \url{https://github.com/akononovicius/lithuanian-parliamentary-election-data}.

In the analysis that follows we consider the parties' vote share distribution
across each polling station. The vote share, $v_{ij}$, is defined
as total number votes cast for the party $V_{ij}$ divided by the
total number of votes cast in that polling station:
\begin{equation}
v_{ij}=\frac{V_{ij}}{\sum_{k=1}^{K}V_{kj}},
\end{equation}
here index $i$ varies over the parties ($K$ is the total number
of parties participating in the election) and index $j$ varies over
the polling stations. We consider the probability and rank-size distributions
of $v_{ij}$ across all of the polling stations during the same parliamentary
election. The probability distribution is estimated using the standard
probability density functions (abbr. PDF). The rank-size distributions
are often used if the data varies significantly in scale, e.g., word
occurrence frequency \cite{Zipf1949}, earthquake magnitudes \cite{Sornette1996JGeoR},
city sizes \cite{Gabaix1999QJE}, cross country income distributions
\cite{Shao2011EPL}. When using this technique the original empirical
data is sorted in descending order. Afterwards the sorted data is
plotted with the rank being the abscissa coordinate and the actual
value being the ordinate coordinate. In our case we sort the parties'
vote shares, $v_{ij}$, for each party $i$ separately to produce
$\tilde{v}_{ik}$, for which 
\begin{equation}
\tilde{v}_{i1}\geq\tilde{v}_{i2}\geq\ldots\geq\tilde{v}_{iM}
\end{equation}
is true. In the above $M$ is a total number of polling stations,
so that index $k$ represents correspond the rank. Note that in this
representation the same polling station may be ranked differently
for diffferent parties (namely $k$ might be different for the same
polling station for different $i$). Evidently these two approaches,
PDFs and rank-size distributions, are inter-related, but using both
of them allows to uncover different statistical patterns.

\subsection{The parliamentary election of 1992\label{subsec:election-1992}}

The parliamentary election of 1992 was held in $2061$ local polling
stations. $17$ parties competed in the parliamentary election, but
only $4$ of them were able to obtain more than $5\%$ of popular
vote. For the sake of simplicity we will use the following abbreviations
for these parties: SK \textendash{} ``S\k{a}j\={u}d¸io koalicija'',
LSDP \textendash{} ``Lietuvos socialdemokrat\k{u} partija'', LKDP
\textendash{} ``Lietuvos Krik¨\v{c}ioni\k{u} demokrat\k{u} partijos,
Lietuvos politini\k{u} kalini\k{u} ir tremtini\k{u} s\k{a}jungos ir
Lietuvos demokrat\k{u} partijos jungtinis s\k{a}ra¨as'', LDDP \textendash{}
``Lietuvos demokratin\.{e} darbo partija''. We have combined the
other $13$ parties to form the ``Other'' party (abbr. O) and considered
the votes cast for the combined ``Other'' party alongside the votes
cast for the $4$ main parties.

\begin{figure}
\begin{centering}
\includegraphics[width=0.9\textwidth]{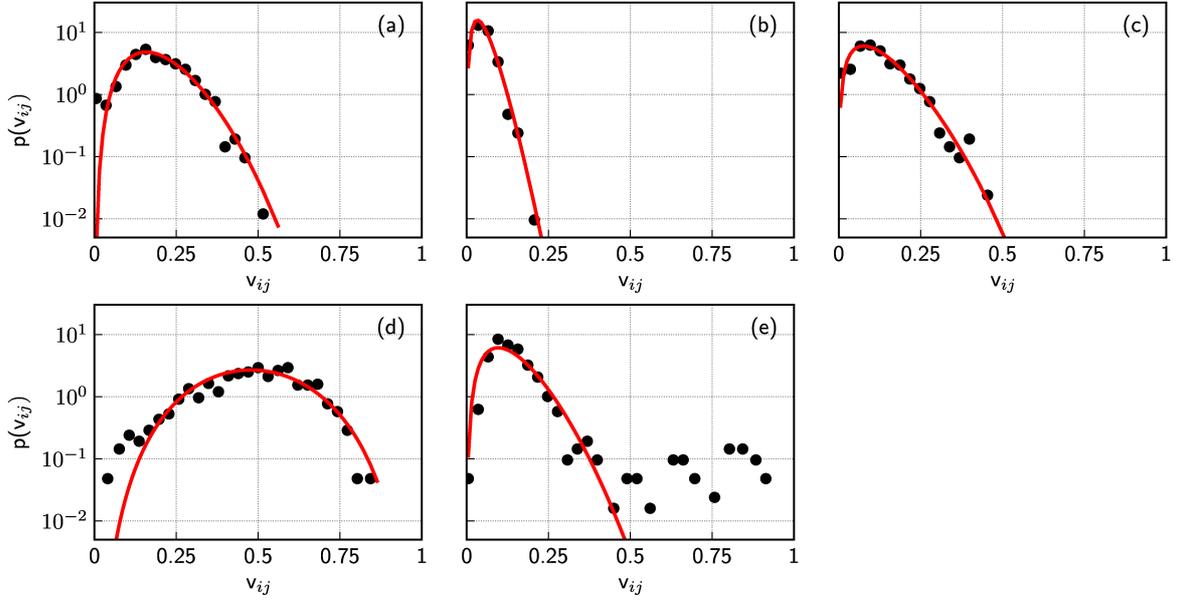}
\par\end{centering}
\caption{The vote share PDF of the most succesful parties during the 1992 election.
The following parties were considered: SK (a), LSDP (b), LKDP (c),
LDDP (d) and O (e). The empirical values are shown as black circles,
while theoretical fits using the Beta distribution are shown as solid
curves. The values of the Beta distribution parameters are given in
Table~\ref{table:params-1992-w}.}

\label{fig:pdf-1992-w}
\end{figure}

\begin{figure}
\begin{centering}
\includegraphics[width=0.9\textwidth]{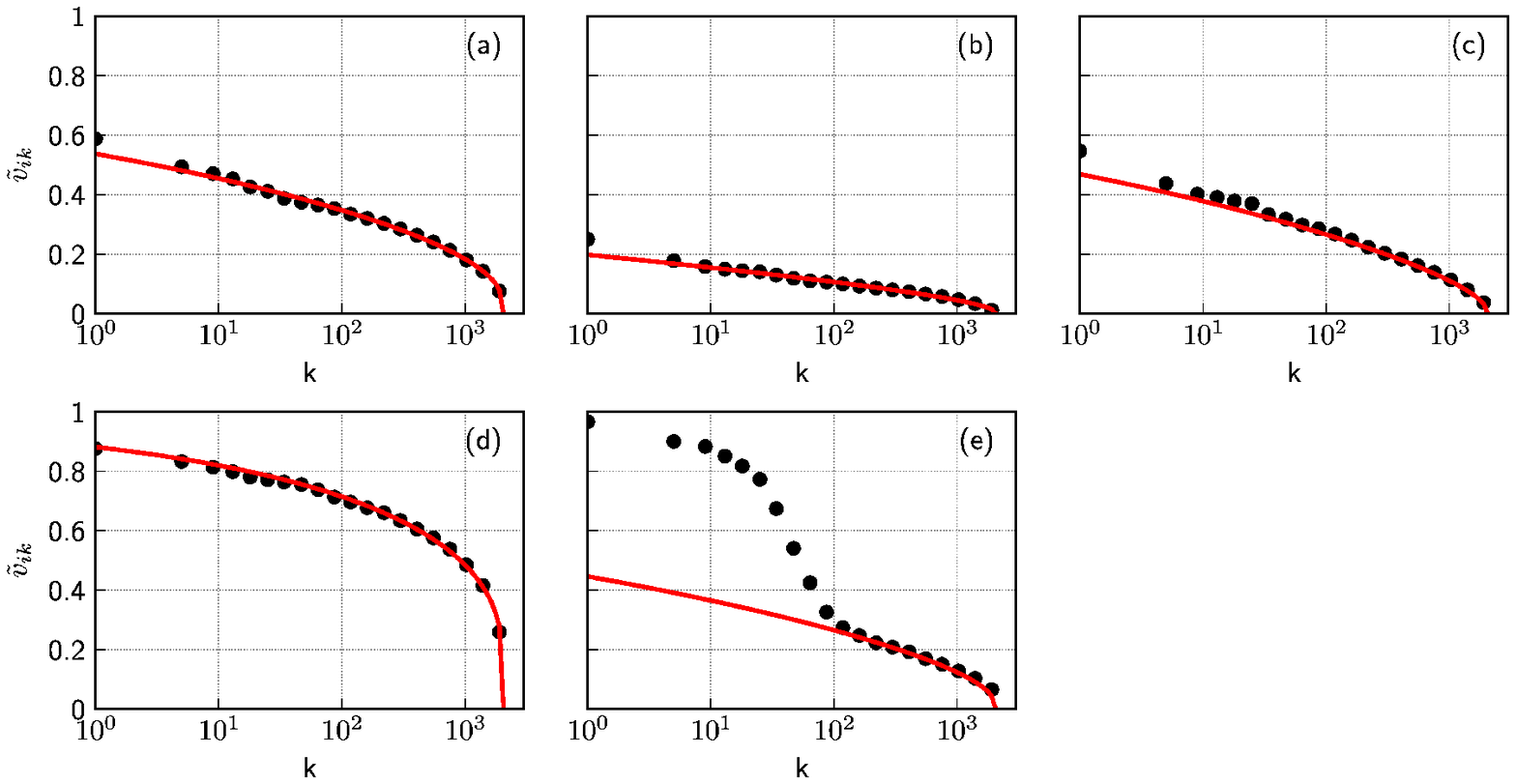}
\par\end{centering}
\caption{The rank-size distribution of the most succesful parties during the
1992 election. The following parties were considered: SK (a), LSDP
(b), LKDP (c), LDDP (d) and O (e). The empirical values are shown
as black circles, while theoretical fits using the Beta distribution
are shown as solid curves. The values of the Beta distribution parameters
are given in Table~\ref{table:params-1992-w}.}

\label{fig:rs-1992-w}
\end{figure}

As you can see from Figs.~\ref{fig:pdf-1992-w} and \ref{fig:rs-1992-w}
as well as Table~\ref{table:params-1992-w} all of the parties with
a notable exception of the ``Other'' party are very well fitted
by assuming that data is distributed according to the Beta distribution,
PDF of which is given by
\begin{equation}
p(v_{ij})=\frac{\Gamma(\alpha)\Gamma(\beta)}{\Gamma(\alpha+\beta)}v_{ij}^{\alpha-1}(1-v_{ij})^{\beta-1}.
\end{equation}
The ``Other'' party stands out, because it includes ``Lietuvos
lenk\k{u} s\k{a}junga'' party (abbr. LLS; en. Association of Poles
in Lithuania). The LLS party had heavily relied on the support of
the ethnic minorities, which were spatially segregated. Namely, the
representatives of ethnic minorities mostly live in larger cities
and Vilnius County. The observed spatial segregation could easily
cause the segregation observed in the voting data.

\begin{table}
\caption{Parameters of the Beta distribution, $\alpha$ and $\beta$, used
to fit the data in Figs.~\ref{fig:pdf-1992-w} and \ref{fig:rs-1992-w}
as well as wellness of fit for the PDFs, $R_{PDF}^{2}$, and the rank-size
distributions, $R_{RS}^{2}$.}
\begin{centering}
\begin{tabular}{|c|c|c|c|c|}
\hline 
Party & $\alpha$ & $\beta$ & $R_{PDF}^{2}$ & $R_{RS}^{2}$\tabularnewline
\hline 
\hline 
SK & $3.9$ & $16.6$ & $0.956$ & $0.994$\tabularnewline
\hline 
LSDP & $2.7$ & $51$ & $0.935$ & $0.953$\tabularnewline
\hline 
LKDP & $2.2$ & $16$ & $0.926$ & $0.995$\tabularnewline
\hline 
LDDP & $5.7$ & $6.1$ & $0.907$ & $0.998$\tabularnewline
\hline 
O & $3$ & $19.4$ & $0.895$ & $0.854$\tabularnewline
\hline 
\end{tabular}
\par\end{centering}
\label{table:params-1992-w}
\end{table}

\begin{figure}
\begin{centering}
\includegraphics[width=0.7\textwidth]{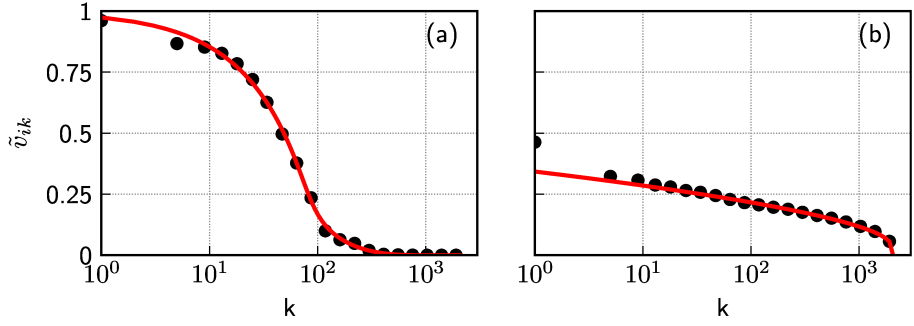}
\par\end{centering}
\caption{The rank-size distributions of the LLS (a) party and the other parties
in the O party (b). The empirical values are shown as black circles,
while solid curves provide fits by assuming that: $v_{ij}\sim0.95\cdot\mathcal{B}e(0.08,10)+0.05\mathcal{\cdot B}e(1.22,1.37)$
($R_{RS}^{2}=0.997$; (a)), $v_{ij}\sim\mathcal{B}e(4.9,35.4)$ ($R_{RS}^{2}=0.992$;
(b)).}

\label{fig:lls-1992-rs}
\end{figure}

In Fig.~\ref{fig:lls-1992-rs} we confirm this intuition by splitting
the LLS party away from the ``Other'' party. After the split the
rank-size distribution of the ``Other'' party is well approximated
by the Beta distribution with parameters $\alpha=4.9$, $\beta=35.4$
($R_{RS}^{2}=0.992$). To provide a good fit for the LLS rank-size
distribution we assume that underlying data is distributed according
to a mixture of the two Beta distributions: one ($95\%$ of points)
with parameters $\alpha_{1}=0.08$, $\beta_{1}=10$ and the other
($5\%$ of points) with parameters $\alpha_{2}=1.22$, $\beta_{2}=1.37$.

The parties' vote share rank-size distributions were previously considered
in \cite{Fenner2016}. Unlike in this paper, Fenner and others assumed
that the parties' vote share is distributed according Weibull distribution,
they have obtained rather good fits for the UK election data. Yet
fits obtained here, assuming Beta distribution, are also rather good.
We believe that Beta distribution is superior for this purpose from
the theoretical point of view. Namely, Beta distribution has reasonable
support, probabilities are defined for $v\in\left[0;1\right]$, while
Weibull distribution needs to be arbitrary truncated, as probabilities
are defined for $v\in\left[0;+\infty\right)$. Interestingly Fenner
and others also use a mixture distribution (of two Weibull distributions)
to fit the UK election data. Similar observations were also made when
studying Brazilian presidential election data \cite{Paz2015}. In
\cite{Ausloos2007EPL} it was noted that multiple different distributions,
Weibull, log-normal and normal, provide good fits for the distribution
of religions' adherents. To discriminate between the possibilities
a deeper theoretical insight is needed.

\subsection{The parliamentary election of 2008}

The parliamentary election of 2008 was held in $2034$ polling stations,
yet we have only $1983$ points in the data set as the data from $51$
polling stations is missing. In this election a slightly smaller number
of parties had participated ($16$), but now $7$ of them were able
to obtain more than $5\%$ of popular vote. For the sake of simplicity
we will use the following abbreviations for them: LSDP \textendash{}
``Lietuvos socialdemokrat\k{u} partija'' (formed by LSDP and LDDP,
which participated in the 1992 election), TS-LKD \textendash{} ``T\.{e}vyn\.{e}s
s\k{a}junga \textendash{} Lietuvos krik¨\v{c}ionys demokratai'' (could
be considered to be a successor of the SK and LKDP, which participated
in the 1992 election), TPP \textendash{} ``Tautos prisik\.{e}limo
partija'', DP \textendash{} ``Koalicija Darbo partija + jaunimas'',
LRLS \textendash{} ``Lietuvos Respublikos liberal\k{u} s\k{a}j\={u}dis'',
TT \textendash{} ``Partija Tvarka ir teisingumas'', LiCS \textendash{}
``Liberal\k{u} ir centro s\k{a}junga''. As previously all other
parties ($9$ of them) were combined to form the ``Other'' party
(abbr. O). We have considered the votes cast for the combined ``Other''
party alongside the votes cast for the $7$ main parties.

\begin{figure}
\begin{centering}
\includegraphics[width=0.9\textwidth]{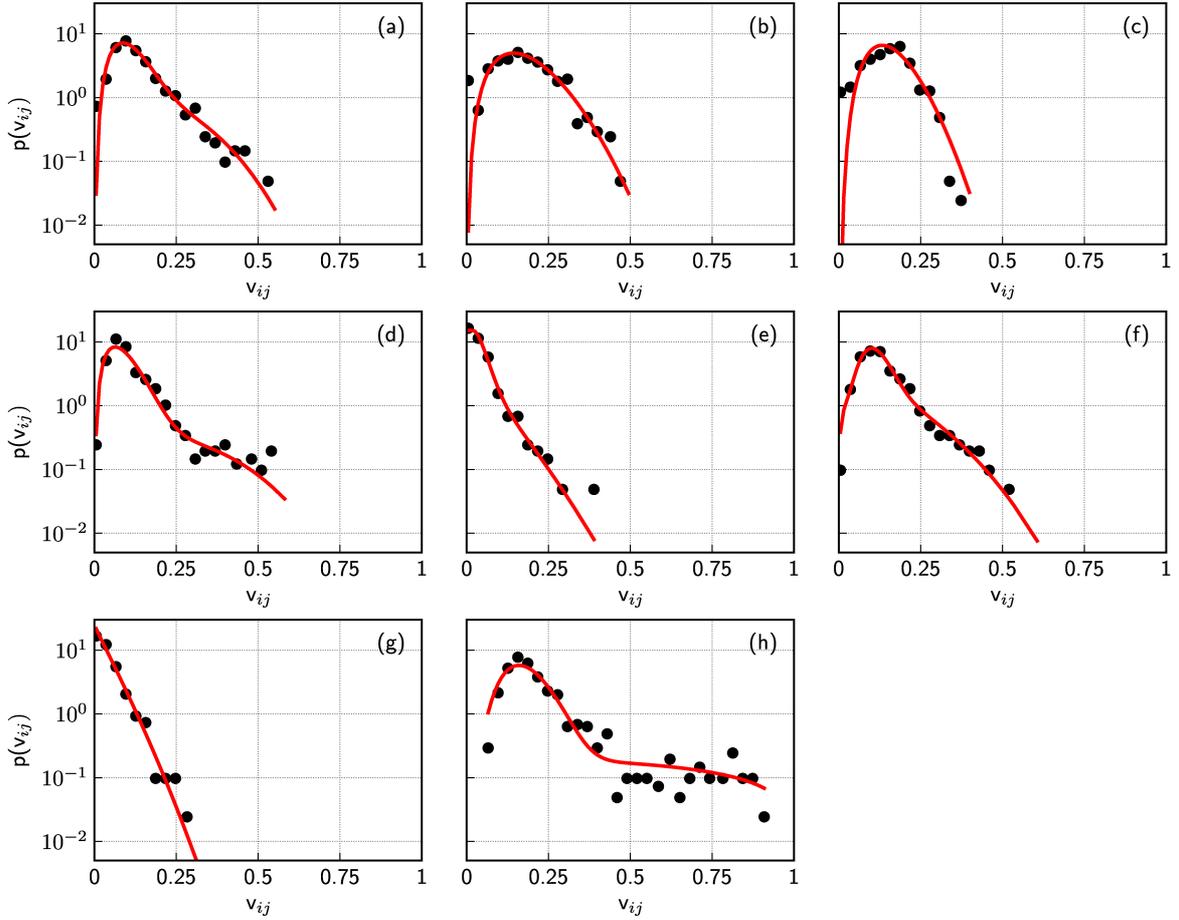}
\par\end{centering}
\caption{The vote share PDFs of the most succesful parties during the 2008
election. The following parties were considered: LSDP (a), TS-LKD
(b), TPP (c), DP (d), LRLS (e), TT (f), LiCS (g) and O (h). The empirical
values are shown as black circles, while theoretical fits using a
mixture of Beta distributions are shown as solid curves (the values
of the parameters are given in Table~\ref{table:params-2008}).}

\label{fig:pdf-2008}
\end{figure}

\begin{figure}
\begin{centering}
\includegraphics[width=0.9\textwidth]{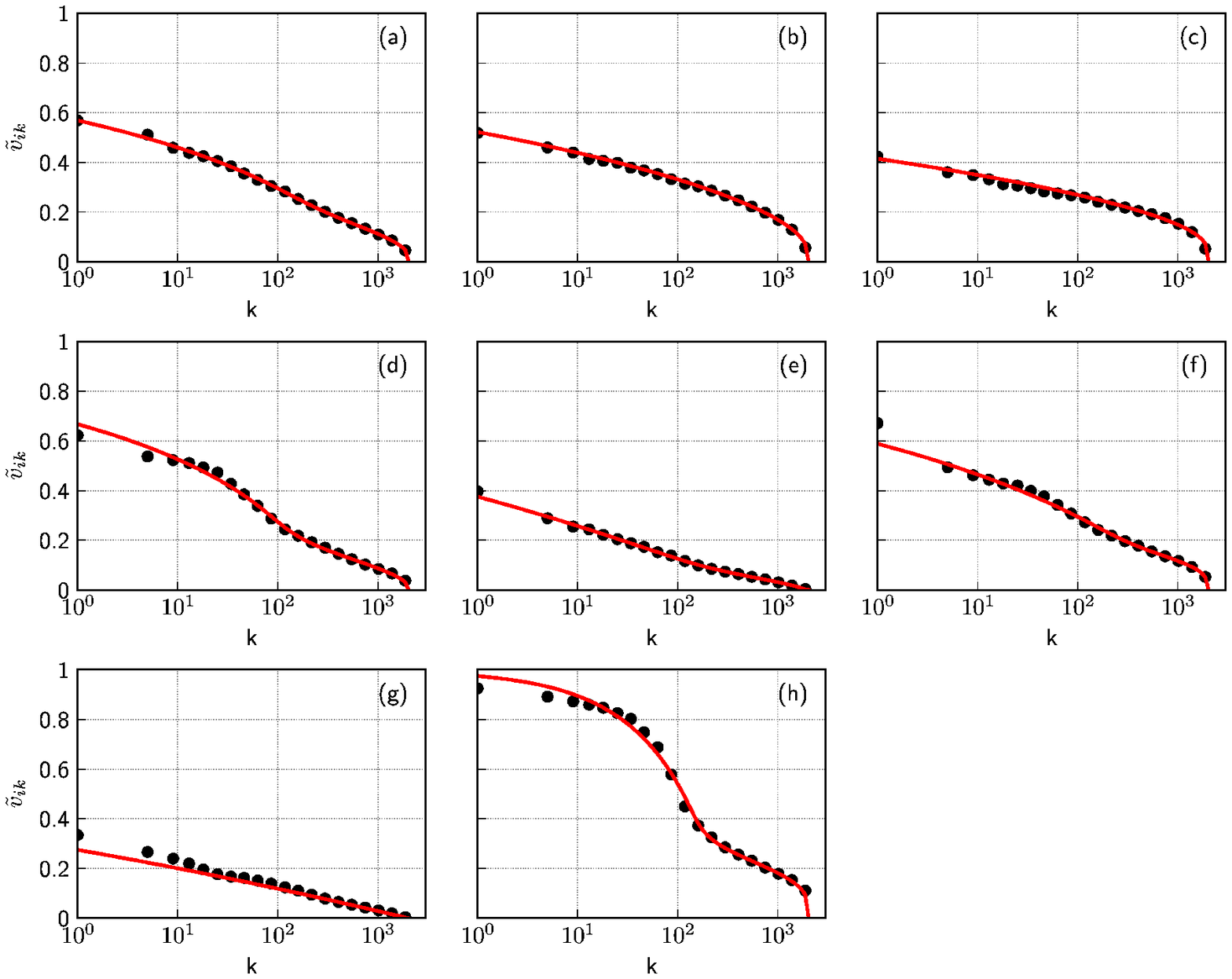}
\par\end{centering}
\caption{The rank-size distributions of the most succesful parties during the
2008 election. The following parties were considered: LSDP (a), TS-LKD
(b), TPP (c), DP (d), LRLS (e), TT (f), LiCS (g) and O (h). The empirical
values are shown as black circles, while theoretical fits using a
mixture of Beta distributions are shown as solid curves (the values
of the parameters are given in Table~\ref{table:params-2008}).}

\label{fig:rs-2008}
\end{figure}

As is evident from Figs.~\ref{fig:pdf-2008} and \ref{fig:rs-2008}
in the 2008 election the vote share distributions of the most of the
parties are well fitted by a mixture of two Beta distributions. Although
now there is no clear-cut explanation for this phenomenon, we would
like to conjecture that this observation indicates that other spatial
segregations (e.g., by income) of voters in Lithuania have started
to play an important role. Namely, most of the parties could now be
identified with specific socio-economic classes, e.g., some party
starts favoring higher income voters (gaining the support in cities),
consequently losing the support of poorer voters (losing the support
in rural areas).

\begin{table}
\caption{The parameters of a mixture of Beta distributions, here $c$ is a
weight of $\mathcal{B}e\left(\alpha_{2},\beta_{2}\right)$, used to
fit the data in Figs.~\ref{fig:pdf-2008} and \ref{fig:rs-2008}
as well as wellness of fit for the PDFs, $R_{PDF}^{2}$, and the rank-size
distributions, $R_{RS}^{2}$.}
\begin{centering}
\begin{tabular}{|c|c|c|c|c|c|c|c|}
\hline 
Party & $\alpha_{1}$ & $\beta_{1}$ & $c$ & $\alpha_{2}$ & $\beta_{2}$ & $R_{PDF}^{2}$ & $R_{RS}^{2}$\tabularnewline
\hline 
\hline 
LSDP & $3.9$ & $31.7$ & $0.15$ & $4.3$ & $12.9$ & $0.968$ & $0.999$\tabularnewline
\hline 
TS-LKD & $3.7$ & $16.8$ & $0$ & \textendash{} & \textendash{} & $0.915$ & $0.999$\tabularnewline
\hline 
TPP & $5.1$ & $27.8$ & $0$ & \textendash{} & \textendash{} & $0.884$ & $0.992$\tabularnewline
\hline 
DP & $3$ & $30.3$ & $0.09$ & $3.2$ & $7.7$ & $0.942$ & $0.992$\tabularnewline
\hline 
LRLS & $2.7$ & $67.9$ & $0.54$ & $0.6$ & $12.6$ & $0.986$ & $0.994$\tabularnewline
\hline 
TT & $7.6$ & $59.5$ & $0.42$ & $1.8$ & $9.4$ & $0.987$ & $0.992$\tabularnewline
\hline 
LiCS & $0.98$ & $23.5$ & $0$ & \textendash{} & \textendash{} & $0.955$ & $0.993$\tabularnewline
\hline 
O & $6.6$ & $30.4$ & $0.15$ & $1.2$ & $1.6$ & $0.944$ & $0.995$\tabularnewline
\hline 
\end{tabular}
\par\end{centering}
\label{table:params-2008}
\end{table}

\subsection{The parliamentary election of 2012}

The parliamentary election of 2012 was held in $2017$ polling stations
(thus we have $2017$ data points). $18$ parties had participated
in the election, while $7$ of them were able to obtain more than
$5\%$ of popular vote. For the sake of simplicity we will use the
following abbreviations for them: LRLS \textendash{} ``Lietuvos Respublikos
liberal\k{u} s\k{a}j\={u}dis'', DP \textendash{} ``Darbo partija'',
TS-LKD \textendash{} ``T\.{e}vyn\.{e}s s\k{a}junga \textendash{}
Lietuvos krik¨\v{c}ionys demokratai'', DK \textendash{} ``Dr\k{a}sos
kelias politin\.{e} partija'', LLRA \textendash{} ``Lietuvos lenk\k{u}
rinkim\k{u} akcija'', LSDP \textendash{} ``Lietuvos socialdemokrat\k{u}
partija'', TT \textendash{} ``Partija Tvarka ir teisingumas''.
Matching abbreviations indicate the same (or mostly the same) parties
as in 2008 election. Once again the less-succesful parties were combined
to form the ``Other'' party (abbr. O). The votes cast for the combined
``Other'' party were analyzed alongside the votes cast for the $7$
main parties.

\begin{figure}
\begin{centering}
\includegraphics[width=0.9\textwidth]{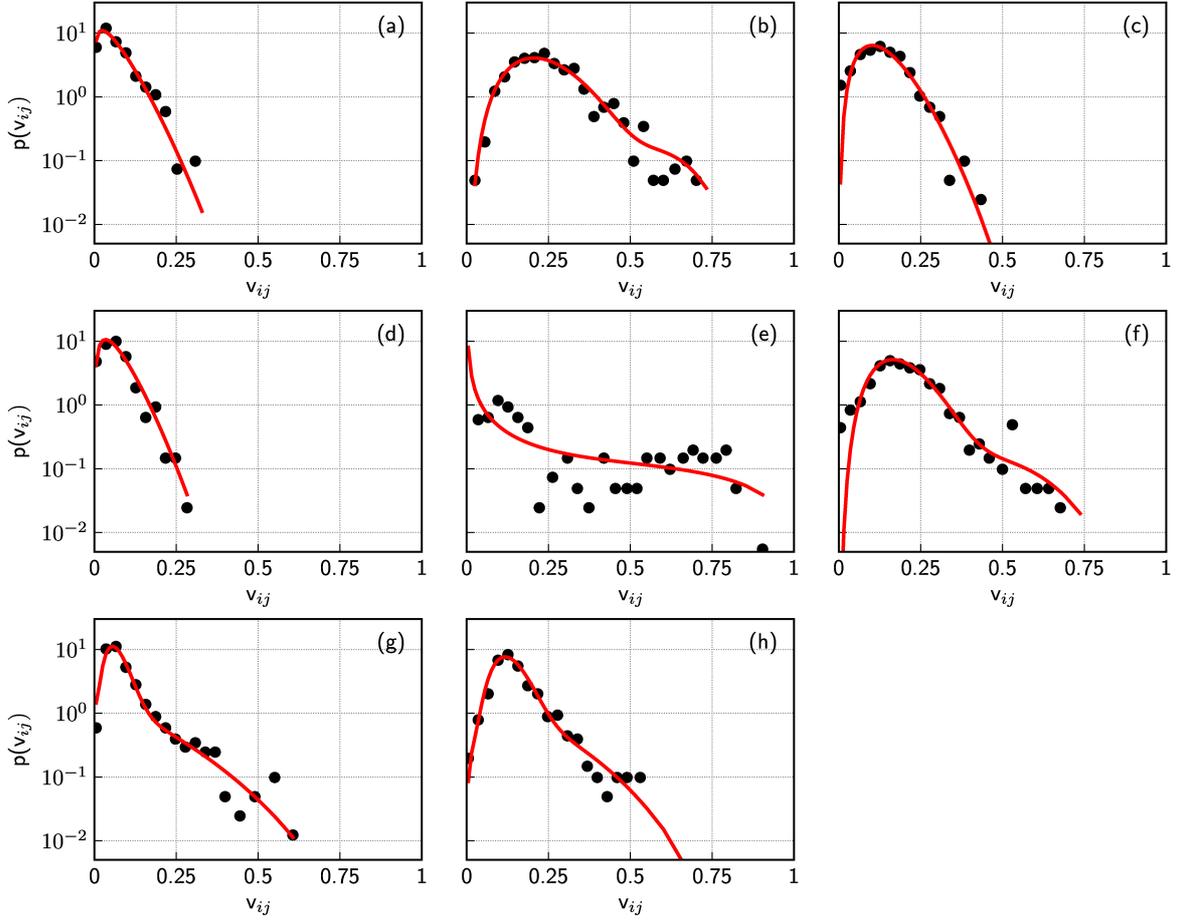}
\par\end{centering}
\caption{The vote share PDFs of the most succesful parties during the 2012
election. The following parties were considered: LRLS (a), DP (b),
TS-LKD (c), DK (d), LLRA (e), LSDP (f), TT (g) and O (h). The empirical
values are shown as black circles, while theoretical fits using a
mixture of Beta distributions are shown as solid curves (the values
of parameters are given in Table~\ref{table:params-2012}).}

\label{fig:pdf-2012}
\end{figure}

\begin{figure}
\begin{centering}
\includegraphics[width=0.9\textwidth]{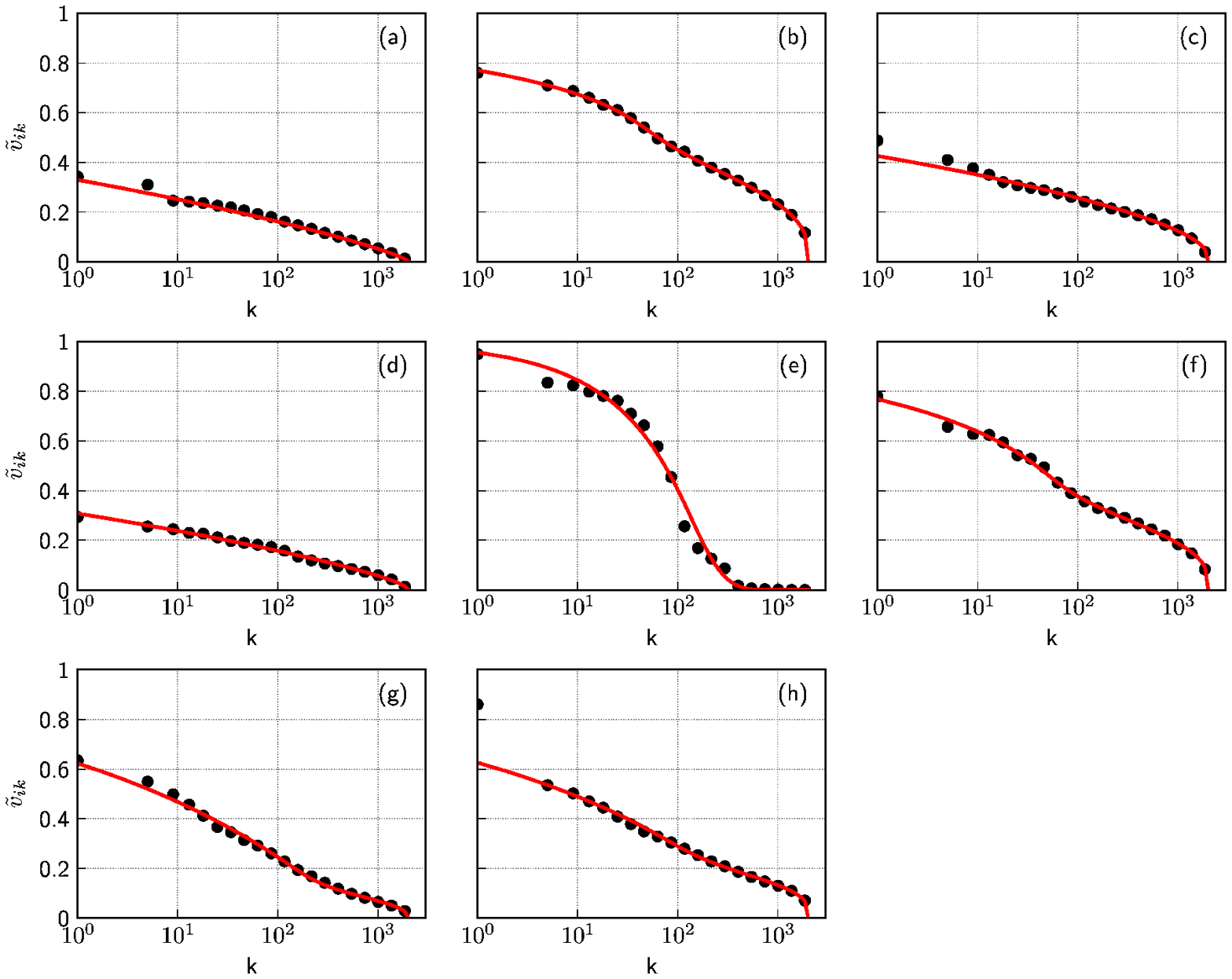}
\par\end{centering}
\caption{The rank-size distributions of the most succesful parties during the
2012 election. The following parties were considered: LRLS (a), DP
(b), TS-LKD (c), DK (d), LLRA (e), LSDP (f), TT (g) and O (h). The
empirical values are shown as black circles, while theoretical fits
using a mixture of Beta distributions are shown as solid curves (the
values of parameters are given in Table~\ref{table:params-2012}).}

\label{fig:rs-2012}
\end{figure}

Once again, as well as in the 2008 parliamentary election data set,
it is evident that the vote share distributions of the most of the
parties in the 2012 parliamentary are also well described by a mixture
of two Beta distributions (see Figs.~\ref{fig:pdf-2012} and \ref{fig:rs-2012}).

\begin{table}
\caption{The parameters of a mixture of Beta distributions, here $c$ is a
weight of $\mathcal{B}e\left(\alpha_{2},\beta_{2}\right)$, used to
fit the data in Figs.~\ref{fig:pdf-2012} and \ref{fig:rs-2012}
as well as wellness of fit for the PDFs, $R_{PDF}^{2}$, and the rank-size
distributions, $R_{RS}^{2}$.}
\begin{centering}
\begin{tabular}{|c|c|c|c|c|c|c|c|}
\hline 
Party & $\alpha_{1}$ & $\beta_{1}$ & $c$ & $\alpha_{2}$ & $\beta_{2}$ & $R_{PDF}^{2}$ & $R_{RS}^{2}$\tabularnewline
\hline 
\hline 
LRLS & $1.5$ & $22$ & $0$ & \textendash{} & \textendash{} & $0.969$ & $0.994$\tabularnewline
\hline 
DP & $4.5$ & $14.5$ & $0.03$ & $15.3$ & $11.1$ & $0.973$ & $0.999$\tabularnewline
\hline 
TS-LKD & $3.4$ & $22$ & $0$ & \textendash{} & \textendash{} & $0.938$ & $0.982$\tabularnewline
\hline 
DK & $1.9$ & $26$ & $0$ & \textendash{} & \textendash{} & $0.946$ & $0.994$\tabularnewline
\hline 
LLRA & $0.06$ & $1.9$ & $0.05$ & $2.1$ & $1.9$ & $0.857$ & $0.990$\tabularnewline
\hline 
LSDP & $5$ & $21.4$ & $0.05$ & $5.5$ & $6.6$ & $0.948$ & $0.997$\tabularnewline
\hline 
TT & $4.6$ & $62$ & $0.29$ & $1.1$ & $6.7$ & $0.963$ & $0.995$\tabularnewline
\hline 
O & $7$ & $45.8$ & $0.23$ & $2$ & $8$ & $0.966$ & $0.970$\tabularnewline
\hline 
\end{tabular}
\par\end{centering}
\label{table:params-2012}
\end{table}

\section{A multi-state agent-based model of the voting behavior\label{sec:abm}}

In this section we propose a simple multi-state agent-based model,
which is to describes the voting behavior within a small non-specific
geographic region covered by a single polling station. Unlike in some
previous approaches \cite{Nowak1990,Axelrod1997JConfRes,Hegselmann2002JASSS,Deffuant2006JASSS,Sobkowicz2012PlosOne,Lilleker2014,Sobkowicz2016PlosOne,Duggins2017JASSS},
our aim is not to incorporate complex ideas from psychology, but to
reproduce the empirical parties' vote share distribution. It is known
that an agent-based herding model proposed by Alan Kirman, in \cite{Kirman1993QJE},
reproduces Beta distribution. So let us start by introducing Kirman's
herding model.

Originally in \cite{Kirman1993QJE} Kirman noted that biologists and
economists observe similar behavioral patterns. Apparently both ants
and people show interest in things which are more popular among their
peers regardless of their objective properties \cite{Bass1969ManSci,Pastels1987Birkhauser1,Pastels1987Birkhauser2,Becker1991JPolitEco,Ishii2012NJP}.
In \cite{Kirman1993QJE} a simple two-state model was proposed to
explain these observations. In the contemporary interpretations of
the Kirman's model the following mathematical form of the one step
transition probabilities is used (see \cite{Alfarano2005CompEco,Alfarano2008Dyncon,Alfi2009EPJB1,Kononovicius2012PhysA}):
\begin{eqnarray}
P(X\rightarrow X+1) & = & \left(N-X\right)\left(\sigma_{1}+hX\right)\Delta t,\label{eq:two-state-1}\\
P\left(X\rightarrow X-1\right) & = & X\left[\sigma_{2}+h\left(N-X\right)\right]\Delta t,\label{eq:two-state-2}
\end{eqnarray}
here $N$ is a total number of agents acting in the system, $X$ is
a total number of agents occupying the first state (consequently there
are $N-X$ agents occupying the second state), $\sigma_{i}$ are the
perceived attractiveness parameters (may differ for different states),
$h$ is a recruitment efficiency parameter and $\Delta t$ is a relatively
short time step.

Let us now show that the two state model produces Beta distribution.
This can be done by using birth\textendash death process formalism,
which is well described in \cite{VanKampen2007NorthHolland}. The
dynamics of $x=\frac{X}{N}$ (let $N$ be large) can be alternatively
described by the following master equation:
\begin{equation}
\frac{\Delta\omega(x,t)}{\Delta t}=N^{2}\left\{ ({\bf E}-1)[\pi^{-}(x)\omega(x,t)]+({\bf E^{-1}}-1)[\pi^{+}(x)\omega(x,t)]\right\} ,
\end{equation}
here $\omega(x,t)$ is time\textendash dependent distribution, $\pi^{\pm}(x)$
are the transition rates per unit of time (defined as $\pi^{\pm}(x)=\frac{p(X\rightarrow X\pm1)}{N^{2}h\Delta t}$),
$\mathbf{E}$ and $\mathbf{E}^{-1}$ are the one step increment and
decrement operators. Let us expand these operators in Taylor series
up to the second order term:
\begin{eqnarray}
{\bf E}[f(x)] & = & f(x+\Delta x)\approx f(x)+\Delta x\frac{\rmd}{\rmd x}f(x)+\frac{\Delta x^{2}}{2}\frac{\rmd^{2}}{\rmd x^{2}}f(x),\\
{\bf E^{-1}}[f(x)] & = & f(x-\Delta x)\approx f(x)-\Delta x\frac{\rmd}{\rmd x}f(x)+\frac{\Delta x^{2}}{2}\frac{\rmd^{2}}{\rmd x^{2}}f(x),
\end{eqnarray}
here $\Delta x=1/N$. Putting these Taylor expansions back into the
master equation as well as taking small time step limit, yields the
following Fokker\textendash Planck equation:
\begin{align}
\frac{\partial}{\partial t}\omega(x,t) & =-\frac{\partial}{\partial x}[A(x)\omega(x,t)]+\frac{1}{2}\frac{\partial^{2}}{\partial x^{2}}[B(x)\omega(x,t)],\\
A(x) & =N\left[\pi^{+}(x)-\pi^{-}(x)\right]=\varepsilon_{1}\left(1-x\right)-\varepsilon_{2}x,\\
B(x) & =\pi^{+}(x)+\pi^{-}(x)=2x(1-x)+\mathcal{O}\left(N^{-1}\right),
\end{align}
where $\varepsilon_{i}=\sigma_{i}/h$. Steady\textendash state distribution
of this Fokker\textendash Planck equation can be obtained by solving:
\begin{equation}
A(x)\omega_{st}(x)-\frac{1}{2}\frac{\partial}{\partial x}[B(x)\omega_{st}(x)]=0.
\end{equation}
In general case the solution of this ordinary differential equation
is given by:
\begin{equation}
\omega_{st}(x)=\frac{C_{0}}{B(x)}\exp\left[2\int^{x}\frac{A(u)}{B(u)}\rmd u\right],
\end{equation}
where $C_{0}$ is normalization constant. In our specific case we
obtain a PDF for the Beta distribution, $\mathcal{B}e\left(\varepsilon_{1},\varepsilon_{2}\right)$,
\begin{equation}
\omega_{st}(x)=\frac{C_{0}}{2}x^{\varepsilon_{1}-1}\left(1-x\right)^{\varepsilon_{2}-1}.
\end{equation}

As in the typical parliamentary election there are more than two competitors,
we need to generalize the model to incorporate more than two states.
From the conservation of the total number of agents $N$, we have:
\begin{equation}
P\left(X_{i}\rightarrow X_{i}\pm1\right)=\sum_{j\neq i}P\left(X_{i}\rightarrow X_{i}\pm1,X_{j}\rightarrow X_{j}\mp1\right).
\end{equation}
Assuming that the right hand side probabilities have the same form
as Eqs.~(\ref{eq:two-state-1}) and (\ref{eq:two-state-2}), we obtain:
\begin{eqnarray}
P(X_{i}\rightarrow X_{i}+1) & = & \sum_{j\neq i}X_{j}\left(\sigma_{ji}+h_{ji}X_{i}\right)\Delta t,\label{eq:two-state-1-2}\\
P\left(X_{i}\rightarrow X_{i}-1\right) & = & X_{i}\sum_{j\neq i}\left[\sigma_{ij}+h_{ij}X_{j}\right]\Delta t,\label{eq:two-state-2-2}
\end{eqnarray}
These one step transition probabilities for $X_{i}$ depend not only
on $X_{i}$ (as in the two-state model), but also on all the other
$X_{j}$ (here $j\ne i$). To circumvent this potentially cumbersome
dependence one needs to assume that $\sigma_{ij}=\sigma_{j}$ and
$h_{ij}=h$ (where $j\neq i$). The first assumption, $\sigma_{ij}=\sigma_{j}$,
means that the perceived attractiveness of any party does not depend
on who is attracted to it. While the second assumption, $h_{ij}=h$,
means that the recruitment mechanism is uniform (symmetric and independent
of interacting agents). Note that these assumptions contrast with
the assumptions underlying the bounded confidence model \cite{Hegselmann2002JASSS,Deffuant2006JASSS}.
Yet these assumptions are needed to ensure that $x_{i}=\frac{X_{i}}{N}$
is distributed according to the Beta distribution. After making these
assumptions we can further simplify the one step transition probabilities:
\begin{eqnarray}
P\left(X_{i}\rightarrow X_{i}+1\right) & = & \left(N-X_{i}\right)\left(\sigma_{i}+hX_{i}\right)\Delta t,\label{eq:np-state-1}\\
P\left(X_{i}\rightarrow X_{i}-1\right) & = & X_{i}\left(\sigma_{-i}+h\left[N-X_{i}\right]\right)\Delta t,\label{eq:np-state-2}
\end{eqnarray}
here $\sigma_{-i}=\sum_{j\neq i}\sigma_{j}$ is the total attractiveness
of all of the competitors of the party $i$. By the analogy with the
two state model it should be evident that
\begin{equation}
x_{i}\sim\mathcal{B}e\left(\varepsilon_{i},\varepsilon_{-i}\right).
\end{equation}
Unlike the two-state model, it seems impossible to provide a useful
general aggregated macroscopic description, using the Fokker-Planck
equation or a set of stochastic differential equations, of the generalized
$M$-state model. In \cite{Kononovicius2013EPL} a three-state model
was considered and given an aggregated macroscopic description, by
a system of two stochastic differential equations, yet it was possible
only under specific conditions.

The one step transition probabilities, while describe agent's behavior,
are still aggregate description of individual agent level dynamics.
So some discussion on what do the Eqs.~(\ref{eq:np-state-1}) and
(\ref{eq:np-state-2}) represent is relevant. Selecting one random
agent, per time step, and setting his switching probability to $\varepsilon_{-i}\Delta t_{s}$
gives us idiosyncratic behavior term, $X_{i}\varepsilon_{-i}\Delta t_{s}$.
While selecting another random agent and, if the both agents vote
for the different parties, allowing the first agent to copy the second
agent's voting preference gives us the recruitment term, $X_{i}(N-X_{i})\Delta t_{s}$.
This description of agent level dynamics could be further generalized
to allow the model to be run on the randomly generated networks \cite{Alfarano2009Dyncon,Kononovicius2014EPJB}.
This agent-based algorithm might be seen to be a special case of the
well known Voter model \cite{Clifford1973,Liggett1999,Galam2002PhysA,Castellano2009RevModPhys,FernandezGarcia2014PRL,Sano2017}.

\section{The modeling of the parliamentary election\label{sec:apply-abm}}

Now let us apply the proposed model to reproduce the empirical vote
share PDFs and rank-size distributions, which were observed during
the 1992 parliamentary election. Here we consider only the simplest
case by ignoring the ``Other'' party and in this way removing distortions
caused by the ethnic segregation (see the discussion in Section~\ref{subsec:election-1992}).
We do not consider the segregated data, the full data of the 1992
parliamentary election or the data of the 2008 and 2012 parliamentary
elections, as to account for the vote segregation a more sophisticated
approach is needed. Namely, in order to account for the full complexity
of the empirical data additional information, such as the spatial
polling data or the socio-demographic data, would be needed. Although,
in general, one could try to infer the correct partition of the polling
stations, where the vote share distribution of each partion would
be modelled using the proposed model using the same parameter set.

\begin{figure}
\begin{centering}
\includegraphics[width=0.5\textwidth]{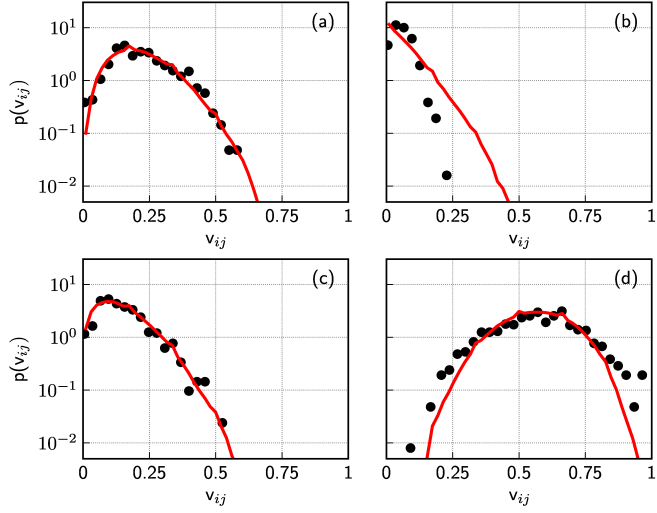}
\par\end{centering}
\caption{The vote share PDFs of the most succesful parties during the 1992
election fitted by the proposed model. Only the following parties
were considered: SK (a), LSDP (b), LKDP (c) and LDDP (d). The empirical
values are shown as black circles, while solid curves represent numerical
results obtained from the proposed model, driven by Eqs.~(\ref{eq:np-state-1})
and (\ref{eq:np-state-2}) (the same model run as in Fig.~\ref{fig:model-rs-1992}).
The parameters of the model are given in Table~\ref{table:params-1992-model}.}

\label{fig:model-pdf-1992}
\end{figure}

\begin{figure}
\begin{centering}
\includegraphics[width=0.5\textwidth]{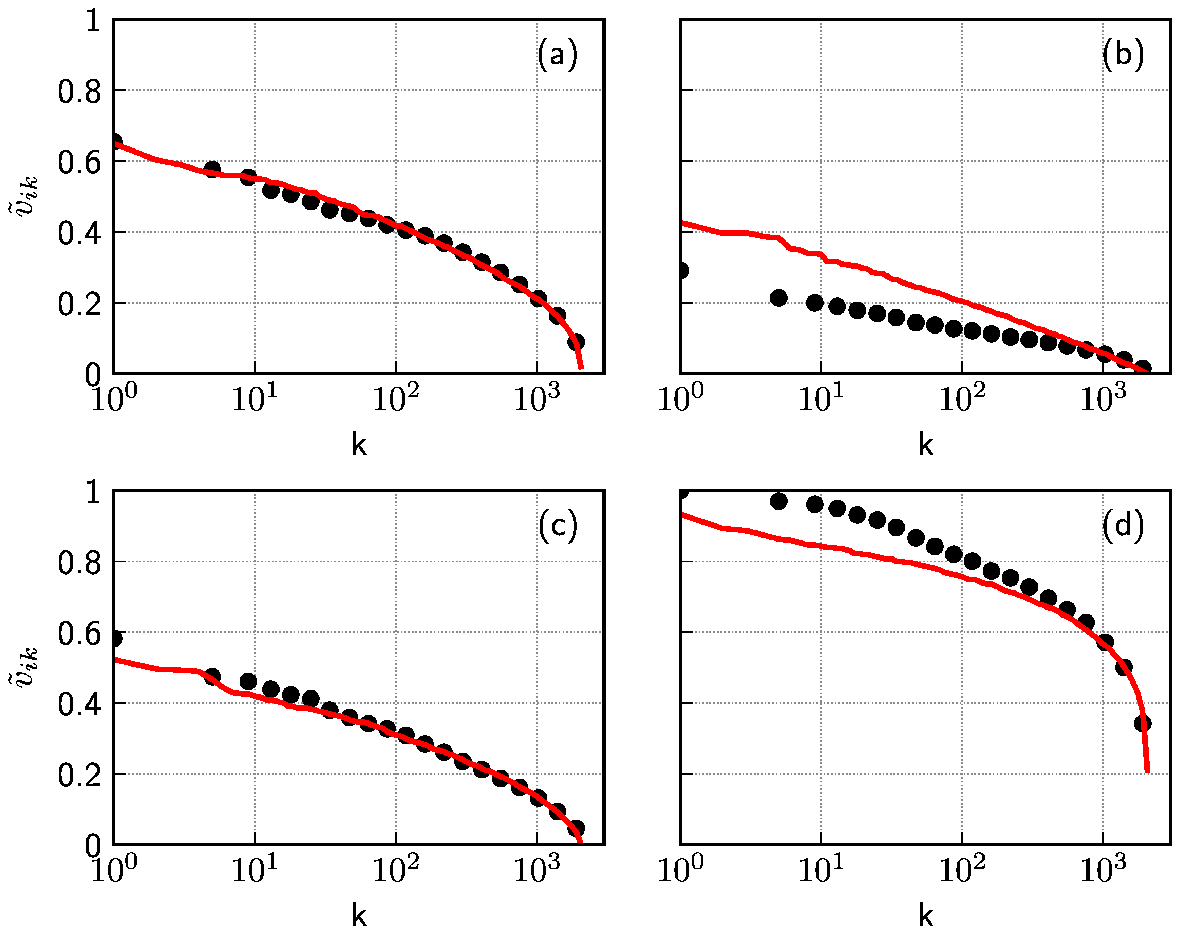}
\par\end{centering}
\caption{The the rank-size distributions of the most succesful parties during
the 1992 election fitted by the proposed model. Only the following
parties were considered: SK (a), LSDP (b), LKDP (c) and LDDP (d).
The empirical values are shown as black circles, while solid curves
represent numerical results obtained from the proposed model, driven
by Eqs.~(\ref{eq:np-state-1}) and (\ref{eq:np-state-2}) (the same
model run as in Fig.~\ref{fig:model-pdf-1992}). The parameters of
the model are given in Table~\ref{table:params-1992-model}.}

\label{fig:model-rs-1992}
\end{figure}

In Figs.~\ref{fig:model-pdf-1992} and \ref{fig:model-rs-1992} we
compare the vote share PDFs and the rank-size distributions numerically
generated by the proposed model, Eqs.~(\ref{eq:np-state-1}) and
(\ref{eq:np-state-2}), and the respective empirical vote share distributions
of the 1992 parliamentary election. Yet we cannot use the previously
empirically estimated Beta distribution parameters, by assuming $\alpha_{i}=\varepsilon_{i}$
and $\beta_{i}=\varepsilon_{-i}$, as model parameters, because an
important model implication, $\sum_{l\neq i}\alpha_{l}=\beta_{i}$,
doesn't hold for the empirical data. Yet one may obtain the parameter
values by fitting the empirical data with the model implication in
mind.

As you can see from Figs.~\ref{fig:model-pdf-1992} and \ref{fig:model-rs-1992}
as well as Table~\ref{table:params-1992-model} the proposed model
excellently fits three of the four parties. While for the LSDP the
fit is not as good as one might expect. Note that model overestimates
the success of LSDP (the solid curve is above black circles for small
$k$ in subfigure (d)) and underestimates the electoral support of
LDDP (the solid curve is below black circles for small $k$ in subfigure
(h)). Thus it is likely that the LSDP had small perceived chance to
win the 1992 election (their aggregated vote share was near $5\%$),
thus voters who would actually consider voting for the LSDP cast their
votes for the other left-wing party, which had better perceived chance
at winning the 1992 election (aggregated vote share of LDDP was above
$40\%$).

\begin{table}
\caption{The parameters of the proposed model used to reproduce the vote share
PDFs and the rank-size distributions of the 1992 election as well
as wellness of fit for the respective distributions.}
\begin{centering}
\begin{tabular}{|c|c|c|c|}
\hline 
Party & $\varepsilon_{i}$ & $R_{PDF}^{2}$ & $R_{RS}^{2}$\tabularnewline
\hline 
\hline 
SK & $3.52$ & $0.951$ & $0.997$\tabularnewline
\hline 
LSDP & $1.13$ & $0.705$ & $0.881$\tabularnewline
\hline 
LKDP & $2.27$ & $0.952$ & $0.998$\tabularnewline
\hline 
LDDP & $8.87$ & $0.904$ & $0.973$\tabularnewline
\hline 
\end{tabular}
\par\end{centering}
\label{table:params-1992-model}
\end{table}

\begin{figure}
\begin{centering}
\includegraphics[width=0.5\textwidth]{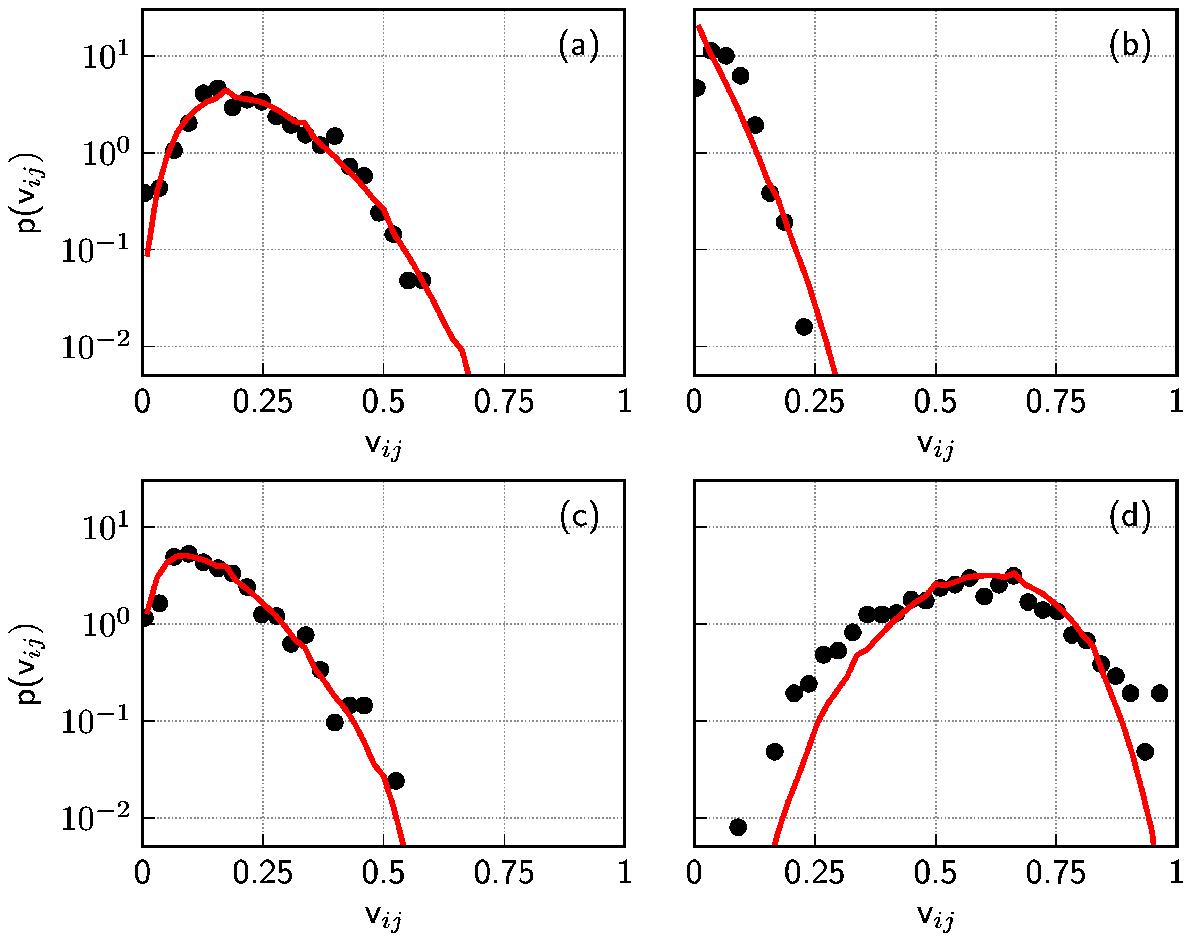}
\par\end{centering}
\caption{The vote share PDFs of the most popular parties during 1992 election
fitted by the asymmetric model. Only the following parties were considered:
SK (a), LSDP (b), LKDP (c) and LDDP (d). The empirical values are
shown as black circles, while solid curves represent data numerically
generated by the proposed model with attractiveness dependent on both
current agent state and the perceived state. Model parameters are
given by Eq.~(\ref{eq:asym-model-params-1992}) (the same model run
as in Fig.~\ref{fig:modelb-rs-1992}).}

\label{fig:modelb-pdf-1992}
\end{figure}

\begin{figure}
\begin{centering}
\includegraphics[width=0.5\textwidth]{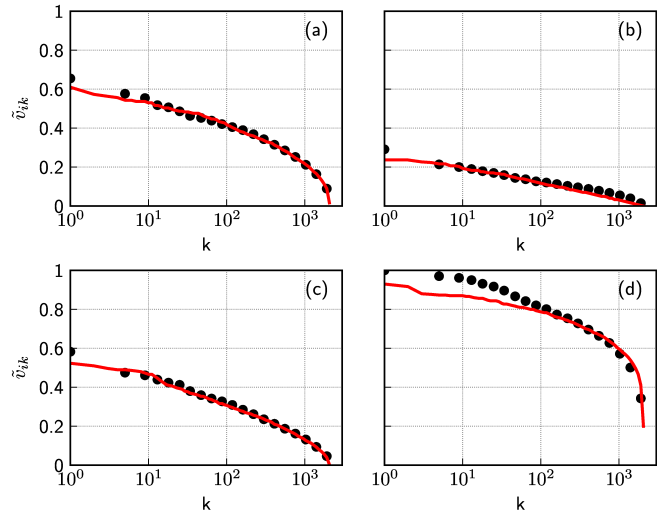}
\par\end{centering}
\caption{The rank-size distributions of the most popular parties during 1992
election fitted by the asymmetric model. Only the following parties
were considered: SK (a), LSDP (b), LKDP (c) and LDDP (d). The empirical
values are shown as black circles, while solid curves represent data
numerically generated by the proposed model with attractiveness dependent
on both current agent state and the perceived state. Model parameters
are given by Eq.~(\ref{eq:asym-model-params-1992}) (the same model
run as in Fig.~\ref{fig:modelb-pdf-1992}).}

\label{fig:modelb-rs-1992}
\end{figure}

We can check this intuition by violating the assumption that the perceived
attractivenes should not depend on the current state of the agent
(agent's currently supported party), namely instead of $\varepsilon_{i}$
we now have $\varepsilon_{ji}$. Previously this assumption was needed
to ensure that vote share is distributed according to the Beta distribution.
Let us introduce a single exception, if $j$ corresponds to LSDP and
$i$ corresponds to LDDP, then $\varepsilon_{ji}$ can be differ in
value from $\varepsilon_{i}$. This gives us the following matrix
of $\varepsilon_{ji}$ values (the numeric indices are assigned according
to Table~\ref{table:params-1992-model}):
\begin{equation}
\boldsymbol{\varepsilon}=\left(\begin{array}{cccc}
0 & 1.13 & 2.27 & 8.87\\
3.52 & 0 & 2.27 & 22\\
3.52 & 1.13 & 0 & 8.87\\
3.52 & 1.13 & 2.27 & 0
\end{array}\right).\label{eq:asym-model-params-1992}
\end{equation}
Note that the diagonal elements of the matrix $\mathbf{\varepsilon}$
are set to zero, as it is not possible to switch to the state the
agent is already in. As you can see in Figs.~\ref{fig:modelb-pdf-1992}
and \ref{fig:modelb-rs-1992}, the fit provided by the model has significantly
improved for the LSDP by just by making this small change.

\section{Conclusions\label{sec:conclusions}}

In this paper we have considered the parties' vote share PDFs and
the rank-size distributions observed during the Lithuanian parliamentary
elections. Namely, we have considered the 1992, 2008 and 2012 parliamentary
elections' data sets. We have determined that the empirical vote share
PDFs and the rank-size distributions are rather well fitted by assuming
that the underlying distribution is the Beta distribution or a mixture
of two Beta distributions. Reviewing literature we have found that
\cite{Ausloos2007EPL,Paz2015,Fenner2016,Sano2017} have reported somewhat
similar results. In \cite{Paz2015,Fenner2016} it was reported that
the empirical data is rather well fitted by a mixture of Weibull distributions.
In \cite{Ausloos2007EPL} it was noted that multiple different distributions,
Weibull, log-normal and normal, provide good fits for the distribution
of religions' adherents. We argue that the Beta distribution is more
suitable as it has correct support (probabilities are defined for
$v\in[0;1]$; although the other distributions could be arbitrary
truncated) and it arises from a simple easily tractable agent-based
model. From our empirical analysis it follows that the mixture of
Beta distributions is needed to fit the data if there is underlying
spatial segregation of the electorate. In \cite{Sano2017} it is also
reported that the empirical data is rather well fitted by Beta distribution,
which arises from a noisy Voter model. Yet \cite{Sano2017} did not
observe the vote share segregation pattern.

Having in mind the stark difference between the psychologically motivated
models, such as bounded confidence model \cite{Hegselmann2002JASSS,Deffuant2006JASSS},
we would like to point out that the observed statistical patterns
as well as the applicability of the model could arise due to numerous
unrelated reasons. One of the alternative possibilities would be the
people mobility patterns. In the proposed model a single agent switching
from supporting one party to supporting another party, could also
represent one agent moving away from the modeled geographic location,
due to social or economic reasons, and another agent, holding different
political views, moving in. A similar idea was raised in \cite{FernandezGarcia2014PRL}.

In the nearest future we will consider spatial modeling of the Lithuanian
parliamentary elections. Another possible approach, with forecasting
possibility, could be considering a temporal regression model for
the attractiveness parameters of the proposed model, $\varepsilon_{i}$,
as well as the estimation of the agent interaction rates, $h$.

\section*{Acknowledgements}

The author would like to thank prof. Ain\.{e} Ramonait\.{e} for suggesting
the idea to analyze the Lithuanian parliamentary election data as
well as for numerous useful remarks. The author would also like to
acknowledge valuable discussions and general interest in this venture
shown by his colleagues at VU ITPA Julius Ruseckas and Vygintas Gontis.


\end{document}